\documentclass{aa}
\def\ltsim{\mathrel{<\kern-1.0em\lower0.9ex\hbox{$\sim$}}}
\def\gtsim{\mathrel{>\kern-1.0em\lower0.9ex\hbox{$\sim$}}}
\usepackage{graphicx}
\begin{document}

\title{The B3--VLA CSS sample. III: EVN \& MERLIN images at  18 cm}

\bigskip

\author{ D. Dallacasa \inst{1,2} \and C. Fanti \inst{3,2}  \and S. Giacintucci
\inst{1,2} \and C. Stanghellini \inst{4} \and R. Fanti \inst{3,2} \and 
L. Gregorini \inst{3,2} \and M. Vigotti \inst{2} }

\offprints{D. Dallacasa}
\mail{ddallaca@ira.cnr.it}

\institute {Dipartimento di Astronomia, Universit\'a di Bologna,
Via Ranzani 1, I--40127 Bologna, Italy \and Istituto di Radioastronomia del
CNR, via Gobetti 101, I--40129 Bologna, Italy \and Dipartimento di Fisica,
Universit\'a di Bologna, Via Irnerio 46, I--40126 Bologna, Italy \and Istituto 
di Radioastronomia del CNR, CP 141, I--96017 Noto SR, Italy} 

\date{Received {{22 January 2002}} / Accepted {{11 April 2002}  } }

      
      \markboth{B3--VLA CSSs}{Dallacasa ed al.}
      \titlerunning{B3--VLA CSSs:EVN\&MERLIN}
      \authorrunning{Dallacasa et al.}

\abstract{
EVN and MERLIN observations at 18 cm are presented for 18 Compact 
Steep--spectrum radio Sources (CSSs) from the B3--VLA CSS sample.
These sources were marginally resolved in previous VLA A-configuration
observations at 4.9 and 8.4 GHz or had peculiar morphologies, two of them
looking like core--jets.
The MERLIN images basically confirm the VLA structures at 8.4 GHz
while the EVN and/or the combined images reveal several additional
details.  
}
\maketitle

\section {Introduction}
\label{intro}
This paper is the third in a series, aimed at studying in detail the
morphology of a new sample of  CSSs ({\bf C}ompact {\bf S}teep--spectrum
{\bf S}ources) \& GPSs ({\bf G}Hz {\bf P}eaked--spectrum {\bf S}ources)
(see O'Dea, \cite{ode2} for the class definition and for a review
of its properties).

The sample (Fanti et al. \cite{fan3}, Paper I) was selected  from the
B3--VLA sample (Vigotti et al. \cite{vig2}) with the purpose of
increasing significantly the existing statistics for sources with
Linear Size  ($LS$)\footnote{$H_0=100\,h$ km\,s$^{-1}$Mpc$^{-1}$,
q$_0=0.5$} in the range 0.4$~h^{-1} \leq LS$(kpc)$ \leq 20~h^{-1}$. 
The scientific motivations for such a project have been illustrated in
the companion papers by Fanti et al. (\cite{fan3}) and by
Dallacasa et al. (\cite{dalla}, Paper II).
The sample consists of 87 CSSs/GPSs and has VLA observations at 1.4
GHz (A and C configurations) and at 4.9 and 8.4 GHz (both A
configuration). A number of sources were not or poorly resolved
even at the highest VLA resolution ($\approx 0.2$ arcsec at 8.4
GHz). For them two VLBI observing projects were undertaken:
VLBA observations, addressed to the most compact sources,
are presented in Paper II. This paper, instead, 
deals with the 18 sources presented in next section.

\section{The source sample and the EVN \& MERLIN observations}
\label{obs}
From the B3-VLA CSS sample we selected the sources which were
slightly resolved ($ 0.2\ltsim \theta \ltsim 1$ arcsec) by the VLA at
8.4 GHz, or showed complex or unclear morphologies on arcsec scales,
requiring further investigation for a better morphological
classification. 

The sample is presented in Table~\ref{tabsample}, which contains the
following information:

\begin{table*}
\vspace{-0.1cm}
\begin{center}
\scriptsize
\begin{tabular}{lcclrrrrcrl}
\hline
\hline
&&&&&&&&&& \\
~~~Name& Id & $m_R$ & ~~~z &  $LAS_{\rm VLA}$ & LogP$_{0.4\rm{GHz}}$ &
S$_{\rm M}$&S$_{{\rm EVN}}$&$LAS_{{\rm E/M}}$&$LLS_{\rm E/M}$ & Morphology \\
& & & &  (arcsec) &  (W/Hz\hspace{2mm} $h^{-2}$) & (mJy)& (mJy) & (arcsec)&
(kpc\hspace{2mm} $h^{-1}$) & \\
~~~~~(1)&(2)&(3)&~~(4)&(5)~~~~~&(6)~~~~~~~&(7)&(8)&(9)&(10)&(11) \\

\hline
\hline
0039+391~~  & G &      & 1.01  & 0.34~~~       & 27.19~~~~~    & 251& 156& 0.38      &       1.6 & MSO  \\
0110+401~~  & Q & 20.0 & 1.479 & $\sim 3.9$~~~ & 27.44~~~~~    & 481& 153&
$\sim$2.2 & $\sim$9.4 & ~~? \\
0123+402~~  & G & 23.8 &       & 1.1~~~        & $>$26.4~~~~~ & 224& 161& 1.22      &       5.2 & MSO \\
0140+387~~  & G &      & 2.9 ~K & 0.7~~~     & 28.55~~~~~    & 340& 116& 1.10      &       4.1 & MSO? \\
0255+460~~  & Q & 20.5 & 1.21  & 0.66~~~       & 27.51~~~~~    & 534& 148& 0.75      &       3.2 & MSO \\
&&&&&&&&&& \\
0722+393A   & E &      &       & 0.25~~~       & $>$26.9~~~~~ & 880& 832& 0.30      & $\sim$1.3 & MSO \\
0748+413B   & E &      &       & $\sim 0.4$~~~ & $>$26.4~~~~~ & 167& ---&
---       &---        & ~~? \\
0754+396~~  & G &      & 2.119 & $\sim 2.2$~~~ & 28.06~~~~~    & 426& 139&
$\sim$2.6 &$\sim$10   & ~~? \\
0810+460B   & G & 20.3 & 0.33 ~R& 0.63~~~       & 26.74~~~~~    & 932& 274& 1.03      &       3.0 & MSO \\
0902+416~~  & E &      &       & 0.34~~~       & $>$26.4~~~~~  & 425& 414& 0.33      & $\sim$1.4 & MSO \\
&&&&&&&&&& \\
1027+392~~  & E &      &       & $\sim 1.6$~~~ & $>$26.4~~~~~ & 339&--- &
---       & ---       & ~~? \\
1039+424~~  & E &      &       & $\sim 1.5$~~~ & $>$26.4~~~~~ & 227&   8&
$\sim$1.5 & $\sim$6.4 & ~~? \\
1128+455~~  & G & 18.7 & 0.40  & $\sim 0.9$~~~ & 26.98~~~~~    &1765&1289& $\sim$0.5 & $\sim$1.6 & MSO? \\
1157+460~~  & G & 21.3 & 0.742 & $\sim 0.8$~~~ & 27.29~~~~~    & 961& 763& 0.62      &       2.5 & MSO? \\
1212+380~~  & G & 24.0 & 1.5 ~~K & 0.3~~~        & 27.63~~~~~    & 247& 104& 0.47      &       2.0 & MSO \\
&&&&&&&&&& \\
1241+411~~  & G & 17.7 & 0.259 & $\sim 1.0$~~~ & 25.72~~~~~    & 330& 169& 0.74      &       1.9 & MSO \\
2301+443~~  & G &      & 1.7 ~~K & 0.5~~~        & 28.34~~~~~    & 975& 737& 0.50      &       2.1 & MSO \\
2349+410~~  & Q & 19.2 & 2.046 & 1.2~~~        & 27.99~~~~~    & 376& 145& 0.82      &       3.3 & MSO \\
\hline
\hline
\end{tabular}
\end{center}
\caption{The B3--VLA CSSs observed with EVN \& MERLIN}
\label{tabsample}
\end{table*}

\smallskip
\noindent
{\it Column 1} -  Source name;

\smallskip\noindent
{\it Column 2} - Optical identification (Id) from Paper I (G~=~galaxy, Q =  
quasar; E = no known optical counterpart);

\smallskip
\noindent
{\it Column 3}  - R magnitude;

\smallskip
\noindent
{\it Column 4 }  - redshift;  ``K'' and ``R'' indicate that the
redshift is estimated by photometric measurements in the respective
optical band (see Paper I for details);

\smallskip
\noindent
{\it Column 5}  - VLA Largest Angular Size,  $LAS$, (arcsec) from Paper I;

\smallskip
\noindent
{\it Column 6}
- log P$_{\rm 0.4 GHz}$ (P in W/Hz $h^{-2}$); for E sources lower
limits to the observed radio power have been computed for $z =
0.5$ (see also Paper I for a wider discussion);  

\smallskip
\noindent
{\it Column 7} - MERLIN  total flux density (mJy) from integration over
the source image (Sect.~\ref{datar}); 

\smallskip
\noindent
{\it Column 8} - EVN total flux density (mJy) from integration over the source 
image (Sect.~\ref{datar}), when available; 

\smallskip
\noindent
{\it Column 9} - Largest Angular Size, $LAS$, (arcsec) from the present paper,
measured on the most appropriate image, (E/M)  (see Sect.~\ref{s-ima}
and Fig.~\ref{images})

\smallskip
\noindent
{\it Column 10} - Largest Linear Size, $LLS$, (kpc) from data in Col.~9.
For E sources $z=1.05$ has been assumed (see also Paper I) and
$LLS$, preceeded by a ``$\sim$'', can be considered a lower limit.

\smallskip
\noindent
{\it Column 11} - morphology as derived from the present paper: MSO - Medium
Symmetric Object (Fanti et al. \cite{fan1}, Readhead et al. \cite{read});
 a ``?''  marks the cases without classification.

The EVN \& MERLIN observations were carried out simultaneously at the
frequency of 1.66 GHz, from June 2nd to June 4th, 1999. The combined
use of both arrays was dictated by the need of studying the source
structures at both high (several mas) and intermediate (tens of mas)
resolutions for a better description of both the compact and the more
resolved structures. 

\begin{figure*}  
\hspace{3.8cm}(a) \vspace {-1 cm}
\hspace{6.5cm}(b)
\begin{center}
\includegraphics{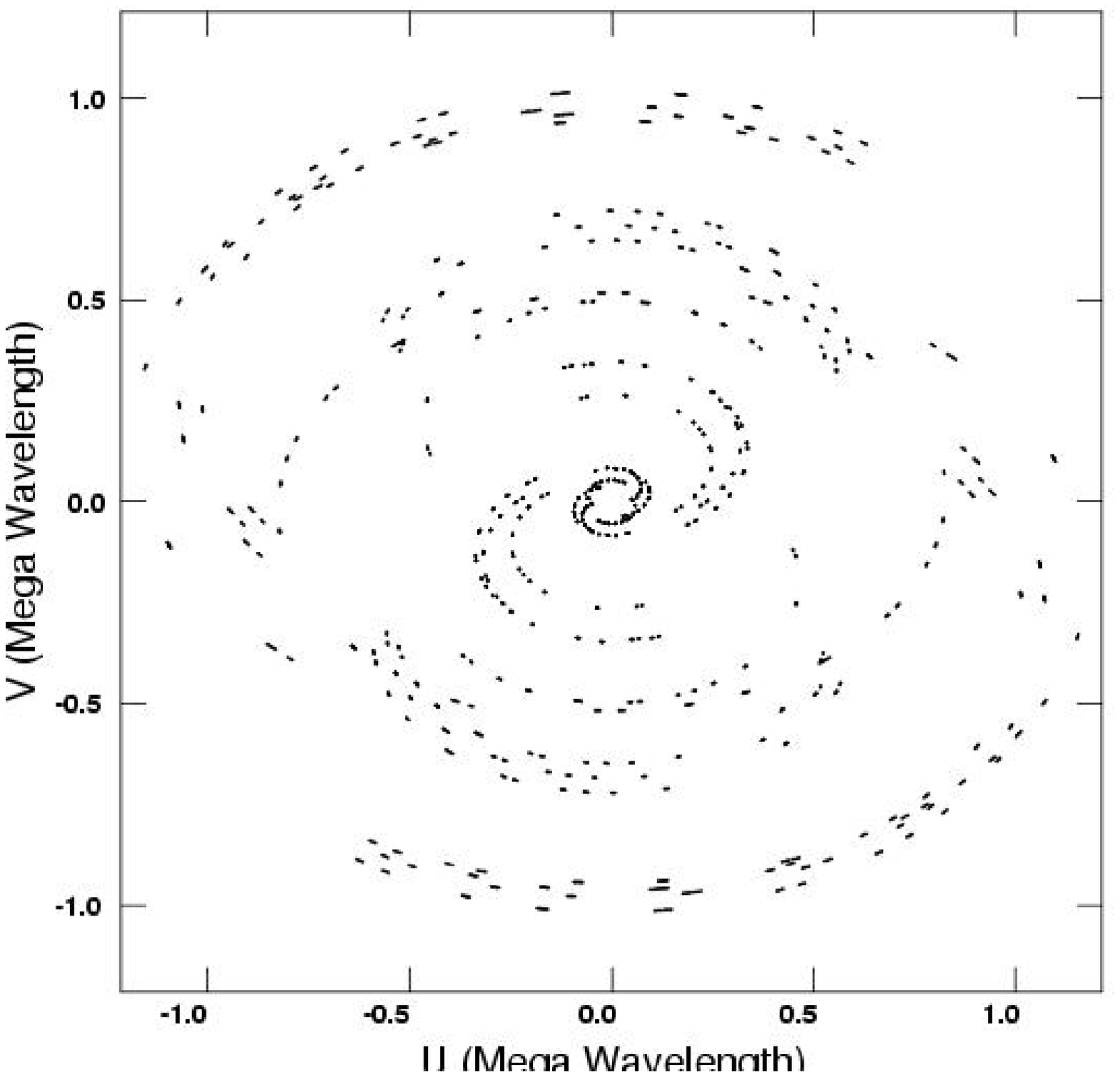}
\includegraphics{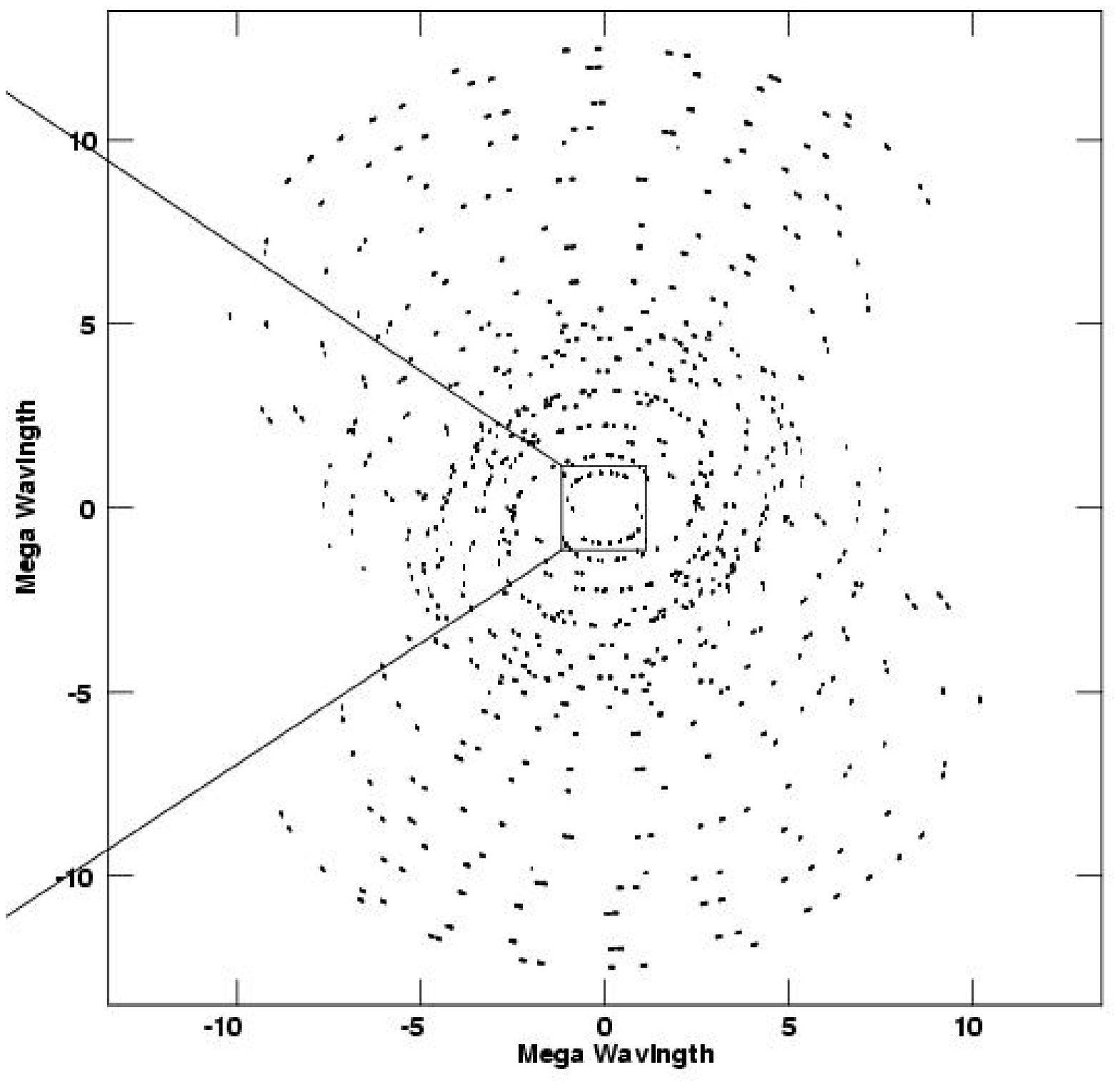}
\end{center}
\vspace{7cm}
\caption{Samples of $uv$ coverage from  MERLIN (a) and EVN (b) }
\label{coperture}
\end{figure*}

The EVN array consisted of the eight telescopes in Effelsberg, Cambridge,
Jodrell Bank (MKII), Medicina, Noto, Onsala (26 m), Westerbork (tied
array) and Toru\'{n},
with projected baselines ranging from $\approx$0.8 M$\lambda$ to
$\approx$13~M$\lambda$. The MERLIN made use
of the six stations of Defford, Cambridge, Knockin, Darnhall, Tabley and MKII
in Jodrell Bank, with $uv$ coverage approximately in the range 0.04--1.2 
M$\lambda$. The baseline
Cambridge--MKII, common to both arrays, was used for consistency checks on
the flux density scales of the two sets of visibility data.

The EVN recorded left-hand circular polarization ($LCP$), with a
total bandwidth of 28 MHz, divided into seven 4--MHz IFs (MkIII
mode~B). The MERLIN recorded both right-hand and left-hand circular
polarization ($RCP$ and $LCP$), with 16 MHz bandwidth for each
polarization hand, at all stations but Cambridge. In fact, due to
limitations in the transmission bandwidth of the MERLIN radio--links,
for this station $LCP$ had to be recorded also on the channel
generally used for $RCP$ in order to be able to use Cambridge {\it
also} as an EVN element  with a bandwidth comparable to that of the
other stations. This resulted in the availability of $LL$ data only in
the MERLIN--only data set for all the baselines to Cambridge.

Every source was observed for a total of about 2 hours.
Target source
observations were interleaved, every 3--4 hours, with observations of the flux
density calibration sources DA193 and OQ208.
In order to obtain a good $uv$ coverage, each source observation was spread
into several scans, each $\approx$ 400 seconds long.
An example of the $uv$ coverage of the two arrays
is given in Fig.~\ref{coperture}. The MERLIN $uv$ coverage is clearly more sparse than that of the 
EVN, due the smaller
number of stations. This however is not critical given the relatively simple
structure of the radio sources at MERLIN resolution and has the merit
of filling the short baseline gap of the EVN, thus ensuring the possibility of
recovering the whole source flux density (Sect.~\ref{datar}) at the
resolution of the combined array.

\section{Data reduction}
\label{datar}
All the data reduction was made using the AIPS software
except for the a--priori MERLIN calibration.  The latter  was performed 
by means of a specific pipeline procedure developed at Jodrell Bank.

The EVN data were correlated with  the Bonn processor at the Max Plank
Institute f\"ur Radioastronomie. Visibility amplitudes were calibrated
using the AIPS standard procedure which makes use of the system
temperature and gain information provided by the  individual
stations. The flux density scale was based on DA193 (S$_{1.7}=$2.03
Jy) and OQ208 (S$_{1.7}=$0.98 Jy) whose flux densities were derived
from the simultaneous MERLIN observations. 

\begin{figure*}
\vspace{6 truecm}
\includegraphics{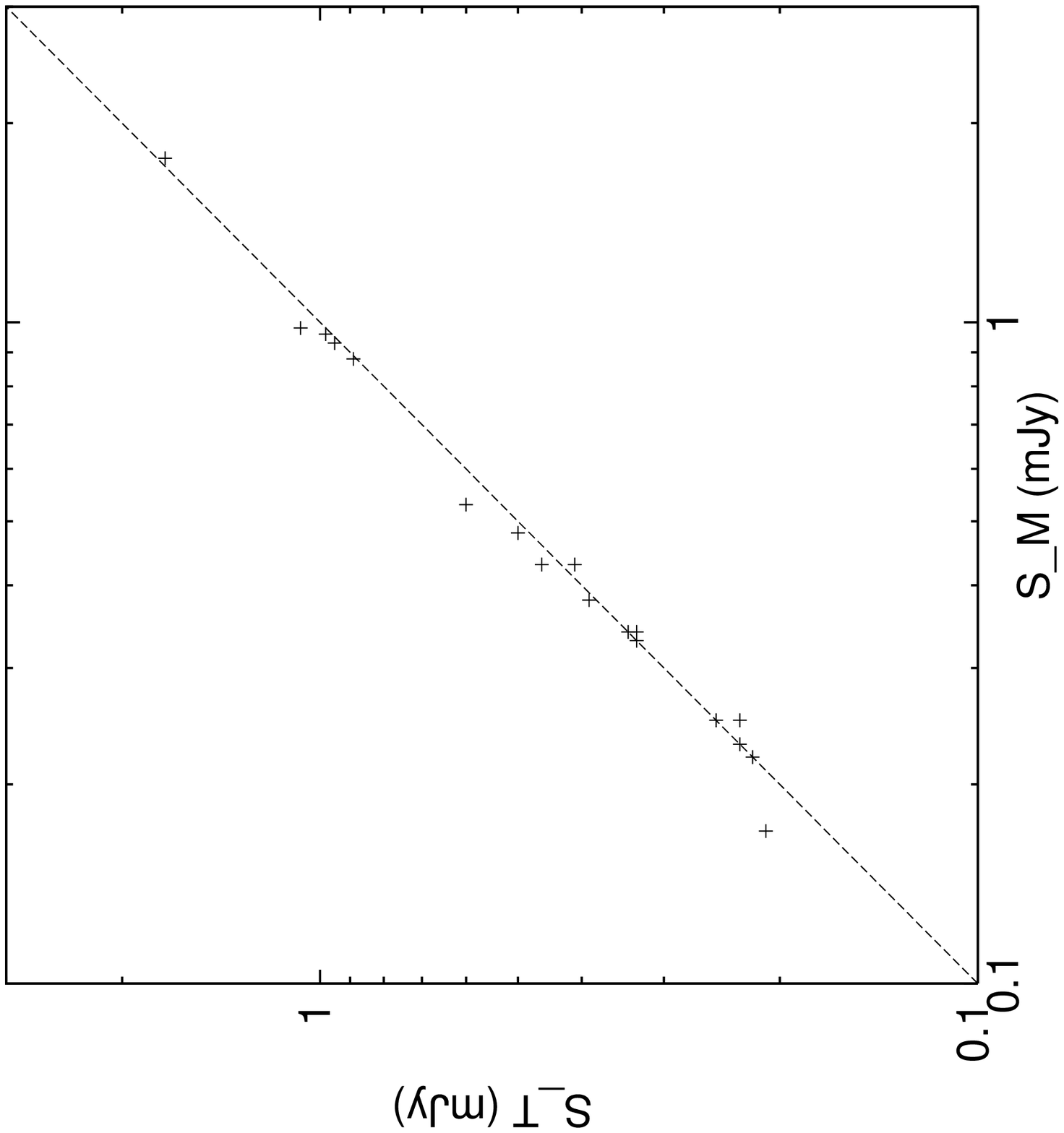}
\includegraphics{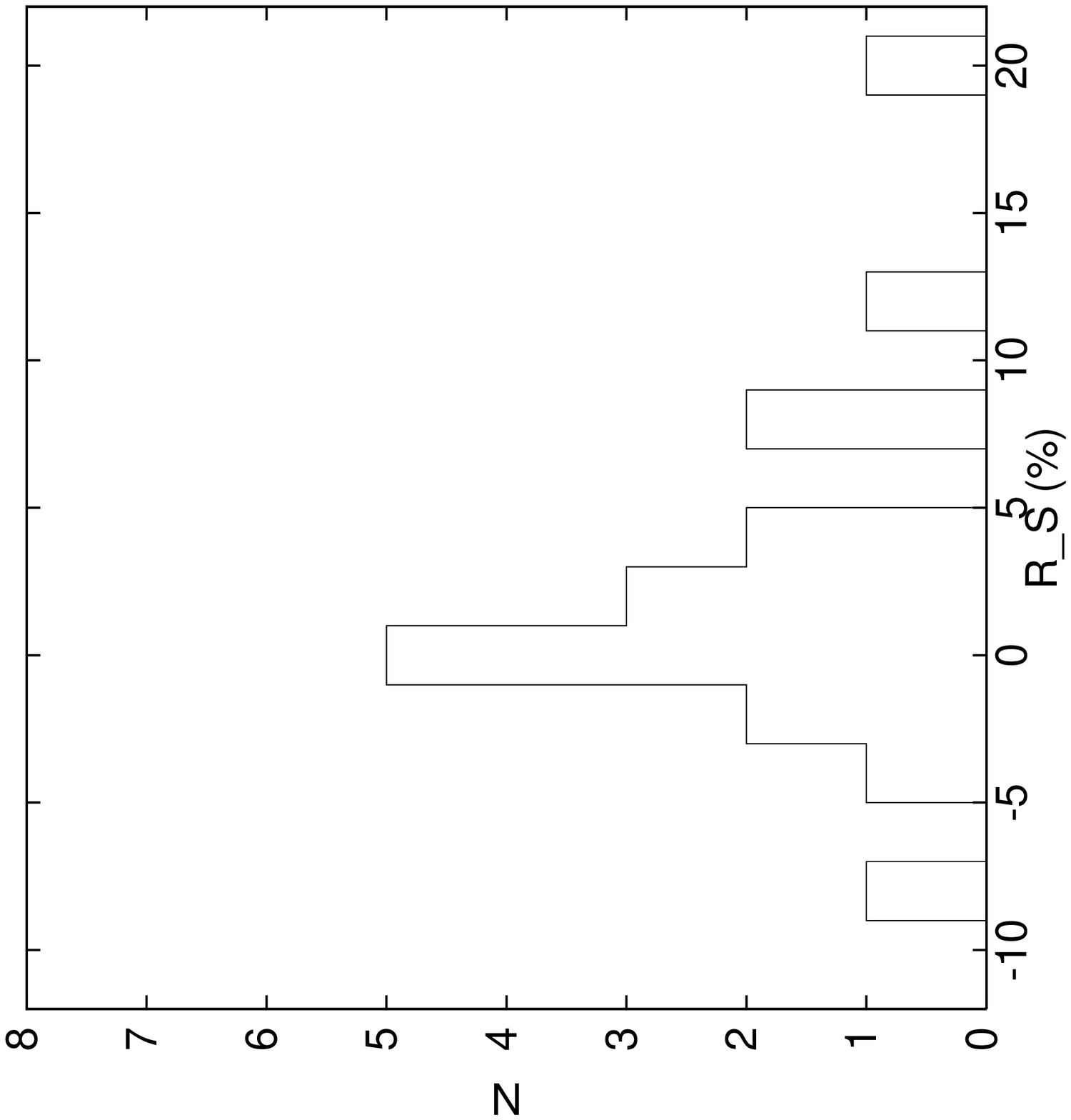}
\vspace{1 truecm}
\caption{ {\it (left)}: S$_{\rm T}$.vs.S$_{\rm M}$; the line has a
slope of one. {\it (right)}: distribution of R$_{\rm S}$; see text for
definitions} 
\label{fluxes}
\end{figure*}

Fringe fitting was then performed by means of the AIPS task FRING. For
some sources the rate of failed solutions exceeded  20--30 \%. This
was not considered satisfactory and two different approaches were
adopted to improve the situation. FRING was run on sub--sets 
of baselines using the same reference antenna in order to force the search for
delay, rate and phase on relatively short baselines. When this did not provide
a significant decrease in the number of failed solutions, the delay,
rate and phase solutions found for the closest well detected target
source were applied. For $0748+413$B and $1027+392$ all the approaches
did not provide significant results, and it was not possible to
obtain an EVN image. For $0110+401, 0123+402, 0754+396$ and
$1212+380$, instead the fraction of failed solutions, occurring mostly
on the longest baselines, is just barely acceptable. The images of
these sources are therefore poorer (see Sect.~\ref{comm}).

We have compared the total flux densities in our MERLIN images ($S_{\rm M}$) with
those at low resolution ($S_{\rm T}$) obtained by interpolating to our observing
frequency the values from the database by Vigotti et al. (\cite{vig}).
The two flux density sets agree satisfactorily with an average ratio
of  $\leq$ 1.02 and a dispersion of 0.06 (r.m.s.). This indicates
that no flux calibration errors nor significant flux density losses
are present in the MERLIN images.
In Fig.~\ref{fluxes} we plot $S_{\rm T}$ vs $S_{\rm M}$ and the histogram of the ratio
$R_{\rm S}=\Delta S/S_{\rm T}$, where $\Delta S=S_{\rm T}-S_{\rm M}$, if positive, represents
the deficiency of MERLIN  flux density with respect to the interpolated total 
flux density. 
Only for two sources (0255+460, 0748+413B) MERLIN appears to have  missed
more than 10\% of the expected flux density (see Sect.~\ref{comm}).
No correlation between $R_{\rm S}$ and $S_{\rm T}$ or $R_{\rm S}$ and $LAS$ is present.

\subsection {Source images}
\label{s-ima}
The radio images were produced initially for MERLIN 
and EVN independently, using the AIPS task IMAGR 
after a number of phase self--calibrations, occasionally ended by a
final amplitude self--calibration. This last step was made with
great care, and for the MERLIN data alone, as the amplitude
self--calibration tends to depress the total flux density when
extended components are poorly sampled in the $uv$ plane.
Afterwards the visibilities from the two arrays were merged and these 
data were self--calibrated again in order to align the phases of the two arrays.

We produced at the end up to three images for each radio source, at {\it low}
(MERLIN--only, $\approx$ 160 mas), {\it intermediate} (EVN \&  MERLIN
$\approx$ 40 mas) and {\it high} (EVN--only, $\approx$ 18 mas)
resolution.
For the two sources $0748+413$B and $1027+392$, whose EVN data were
quite poor, only the MERLIN images are given.

In general the source flux density has been fully recovered in the combined
EVN \& MERLIN images and agrees with the MERLIN total flux density. The ratio 
of these flux densities  has a dispersion of 0.09 (r.m.s.) around the
mean value of unity. Exceptions are mentioned in Sect.~\ref{comm}. 

The actual median  r.m.s. noise level ($\sigma$), measured on the images far
from the sources, is, with a few exceptions,  around 0.2 mJy/beam, not far
from the thermal noise, for all three sets of images.  
The median dynamic range, defined as the ratio of peak brightness to 1
$\sigma$ of the noise, is $\approx$ 1600:1 for MERLIN--only, $\approx$ 300:1 
for EVN \& MERLIN and $\approx$ 100:1 for EVN--only.

\begin{figure*}
\resizebox{\hsize}{!}{
\includegraphics{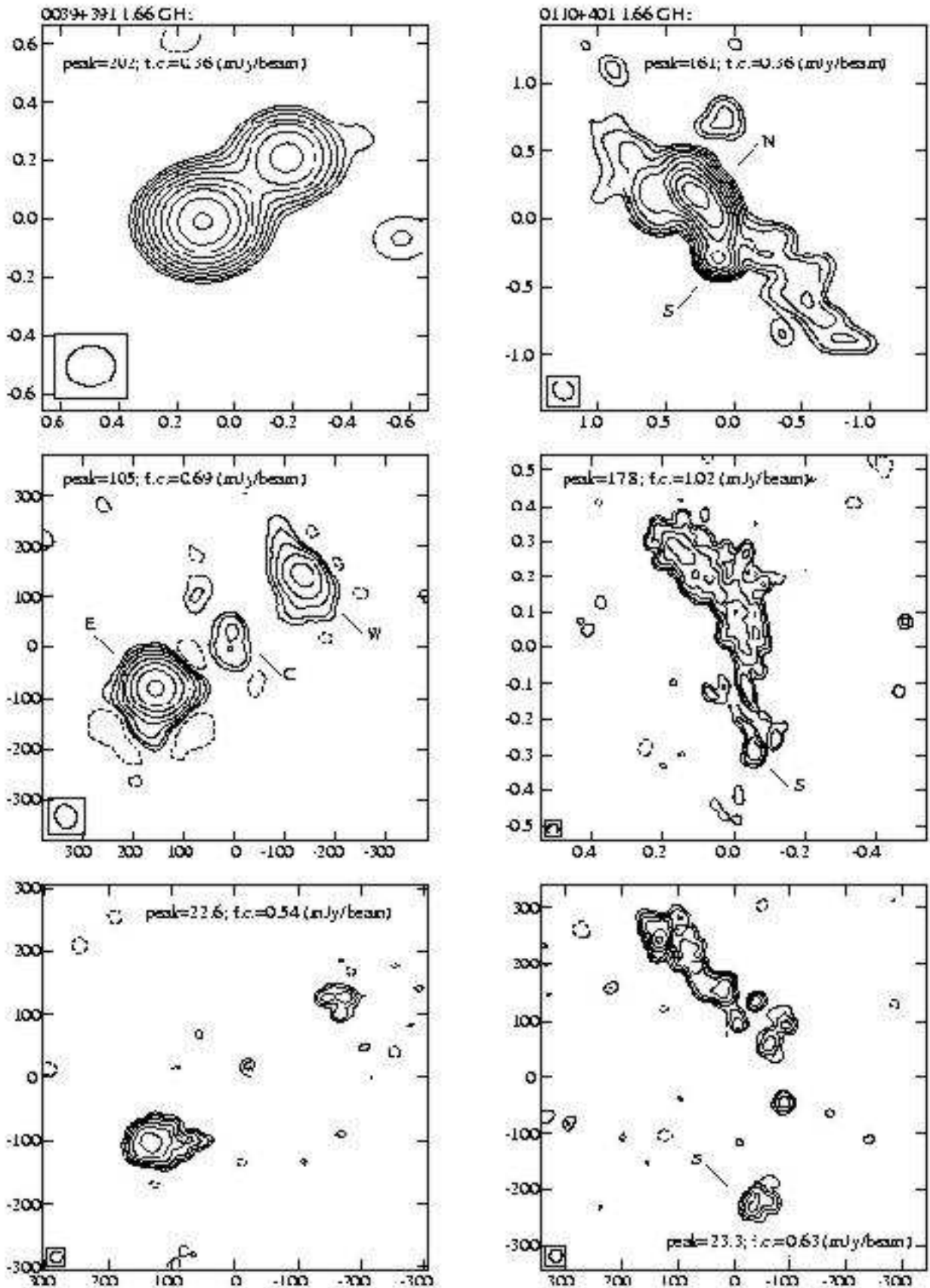}
}
\caption[]{From top to bottom: MERLIN, combined EVN \& MERLIN and EVN
images; the first contour (f.c.) is three times the r.m.s. noise level
on the image; contour levels increase by a factor of 2; the restoring
beam is shown in the bottom left corner in each image.} 
\label{images}
\end{figure*}

All images are shown in Fig.~\ref{images} and Fig.~\ref{mer-ima}. They are
usually displayed in R.A. order; exceptions are due to layout constraints. 
On the plot itself the following information is found:
{\it a)} peak flux density in mJy/beam; {\it b)} first contour ({\it f.c.}) in 
mJy/beam corresponding to the 3 r.m.s.  actual noise of the image; contours 
increase by a factor of~2; {\it c)}~beam  Half Maximum
Width (HMW), represented by the lower left--hand corner ellipse.

Moreover: {\it i)} components are 
labelled according to Table~\ref{comps} on the more convenient image; {\it 
ii)} relative RA and Dec are expressed either in mas (integer) or arcsec 
(decimal); {\it iii)} the coordinate origin corresponds to the center
of each image shown in Fig.~\ref{images} and has no relation with the
actual source position; 

\setcounter{figure}{2}
\begin{figure*}
\resizebox{\hsize}{!}{
\includegraphics{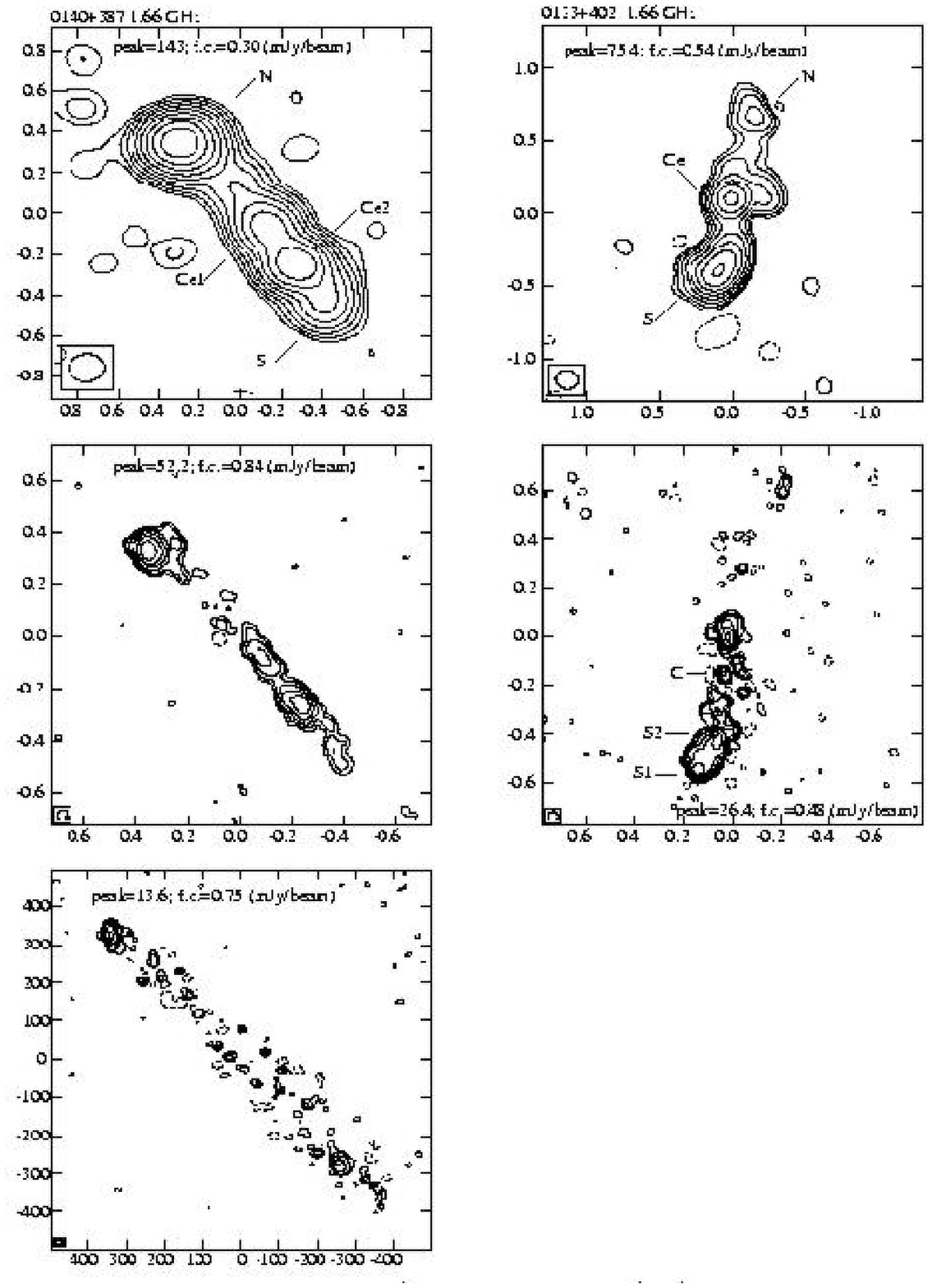}
}
\caption[]{{From top to bottom: MERLIN, combined EVN \& MERLIN and EVN
images (cont.)}}
\end{figure*}
\setcounter{figure}{2}
\begin{figure*}
\resizebox{\hsize}{!}{
\includegraphics{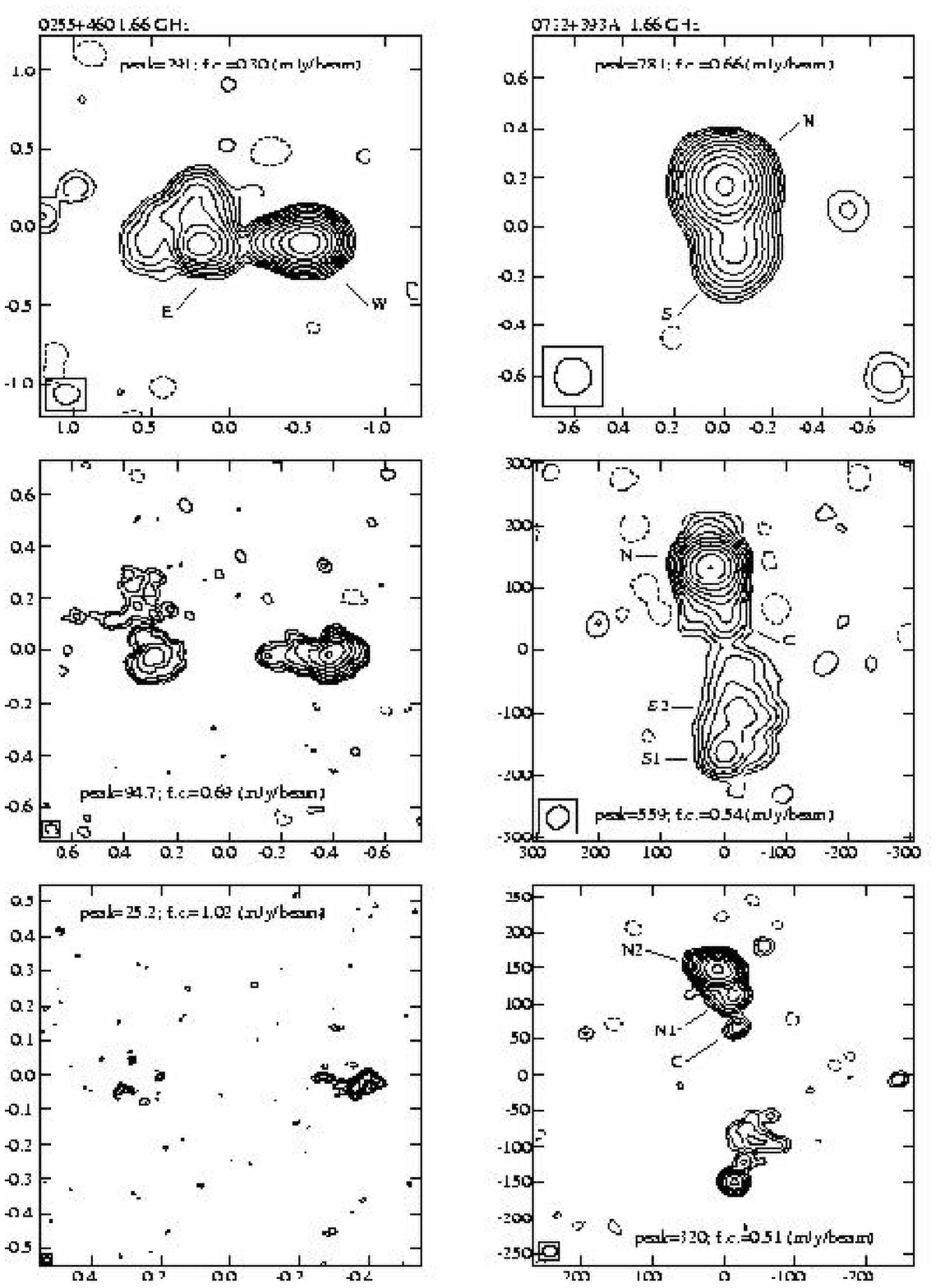}
}
\caption[]{{From top to bottom: MERLIN, combined EVN \& MERLIN and EVN
images (cont.)}}
\end{figure*}
\setcounter{figure}{2}
\begin{figure*}
\resizebox{\hsize}{!}{
\includegraphics{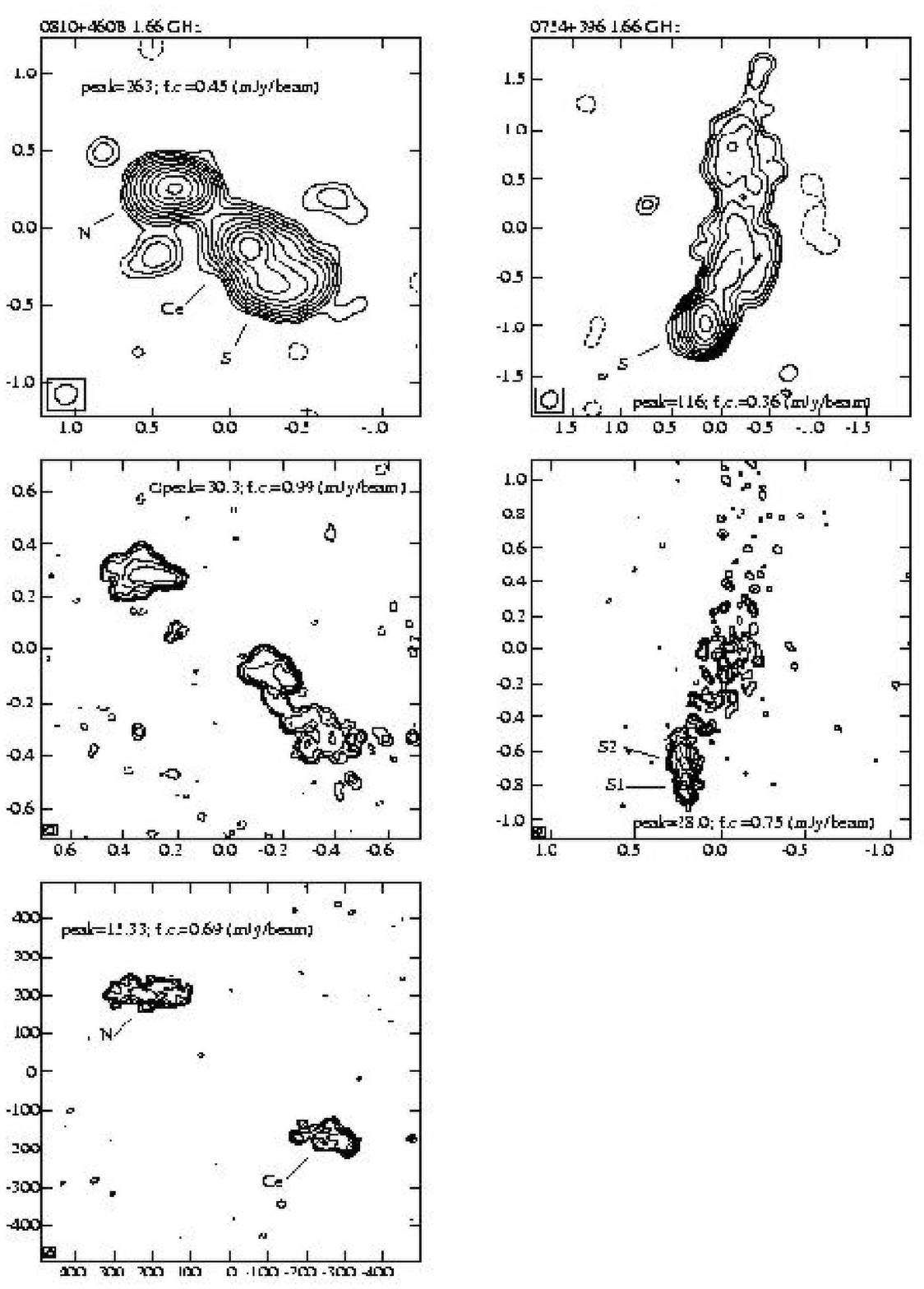}
}
\caption[]{{From top to bottom: MERLIN, combined EVN \& MERLIN and EVN
images (cont.)}}
\end{figure*}
\setcounter{figure}{2}
\begin{figure*}
\resizebox{\hsize}{!}{
\includegraphics{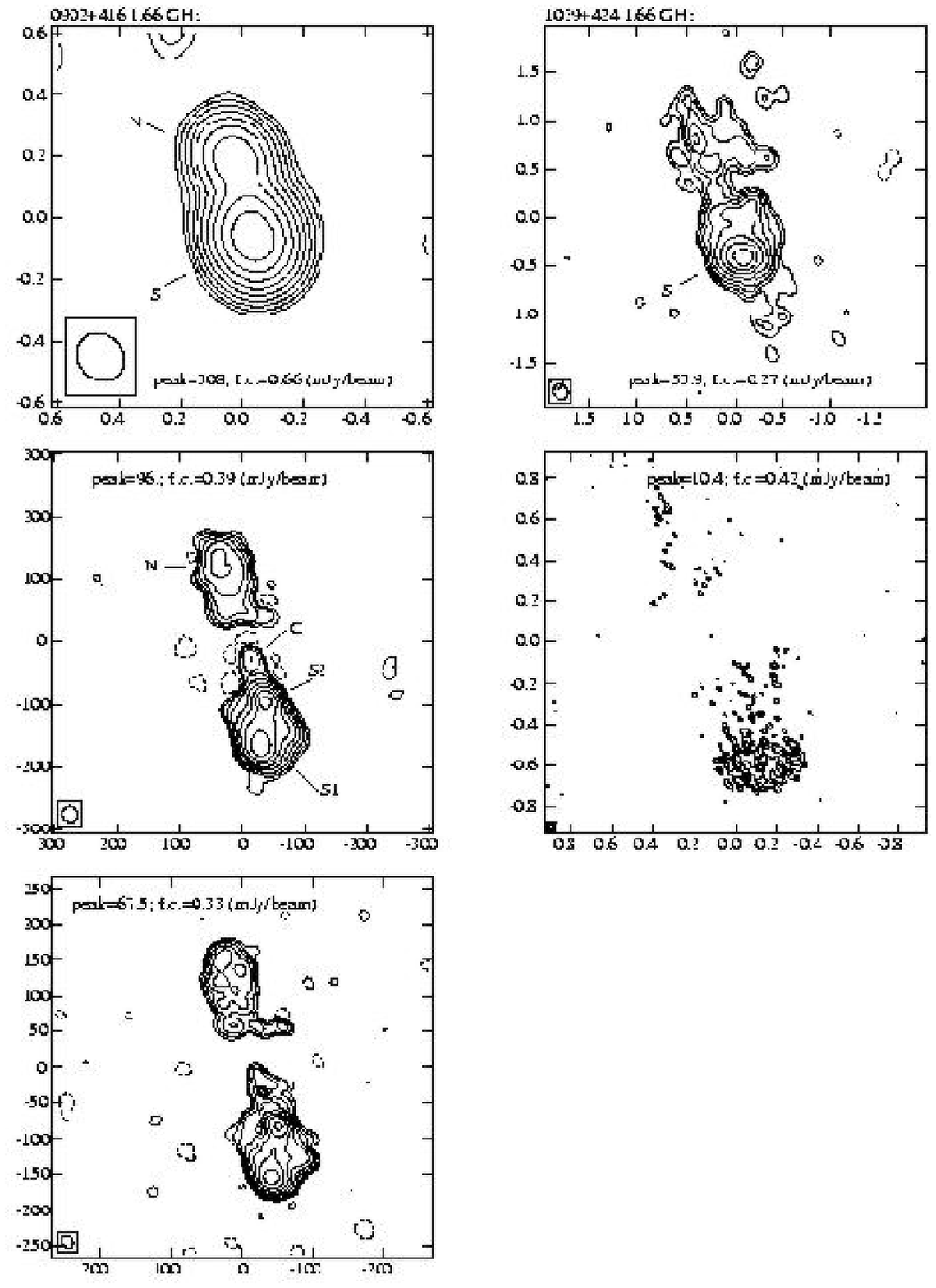}
}
\caption[]{{From top to bottom: MERLIN, combined EVN \& MERLIN and EVN
images (cont.)}}
\end{figure*}
\setcounter{figure}{2}
\begin{figure*}
\resizebox{\hsize}{!}{
\includegraphics{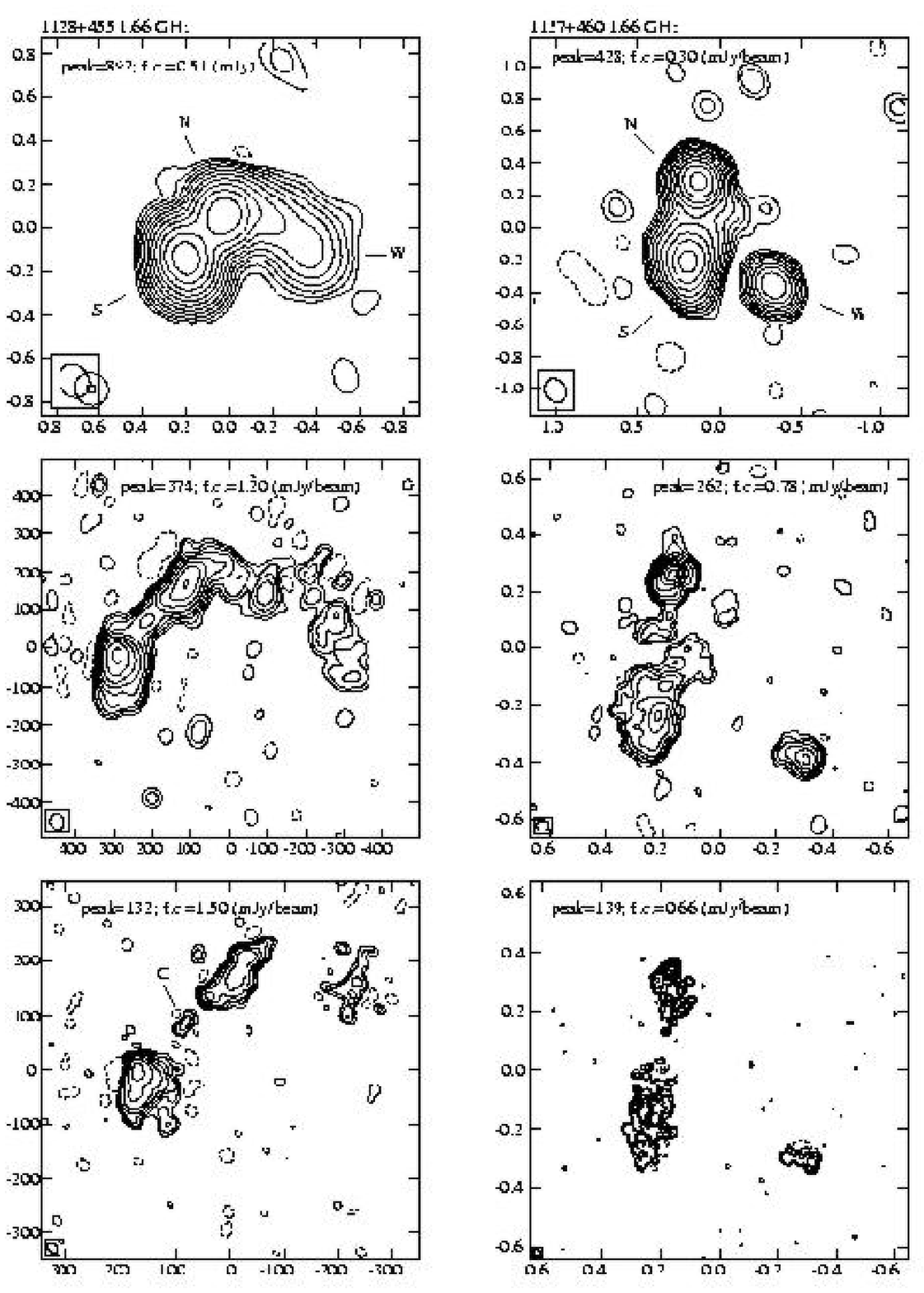}
}
\caption[]{{From top to bottom: MERLIN, combined EVN \& MERLIN and EVN
images (cont.)}}
\end{figure*}
\setcounter{figure}{2}
\begin{figure*}
\resizebox{\hsize}{!}{
\includegraphics{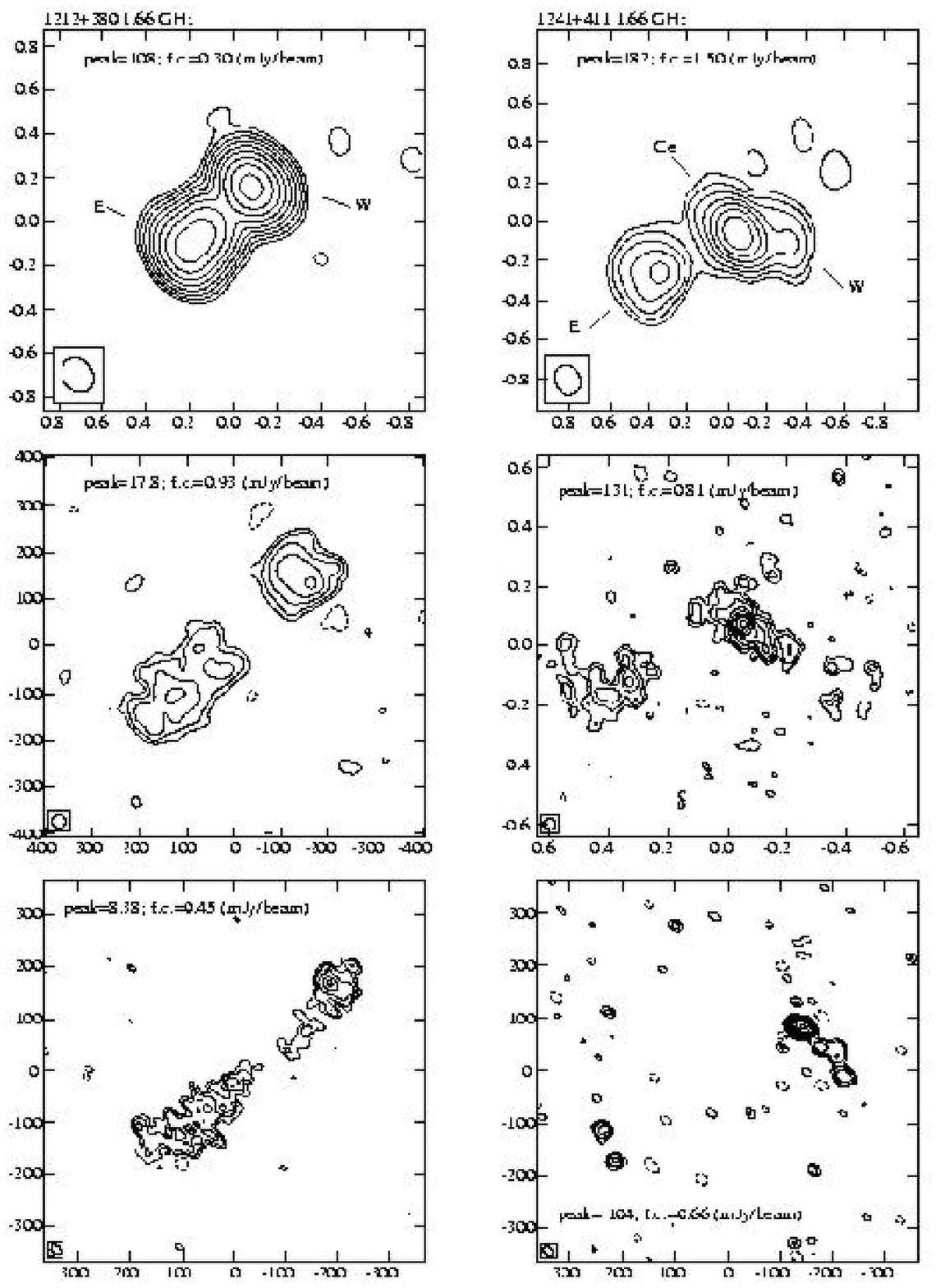}
}
\caption[]{{From top to bottom: MERLIN, combined EVN \& MERLIN and EVN
images (cont.)}}
\end{figure*}
\setcounter{figure}{2}
\begin{figure*}
\resizebox{\hsize}{!}{
\includegraphics{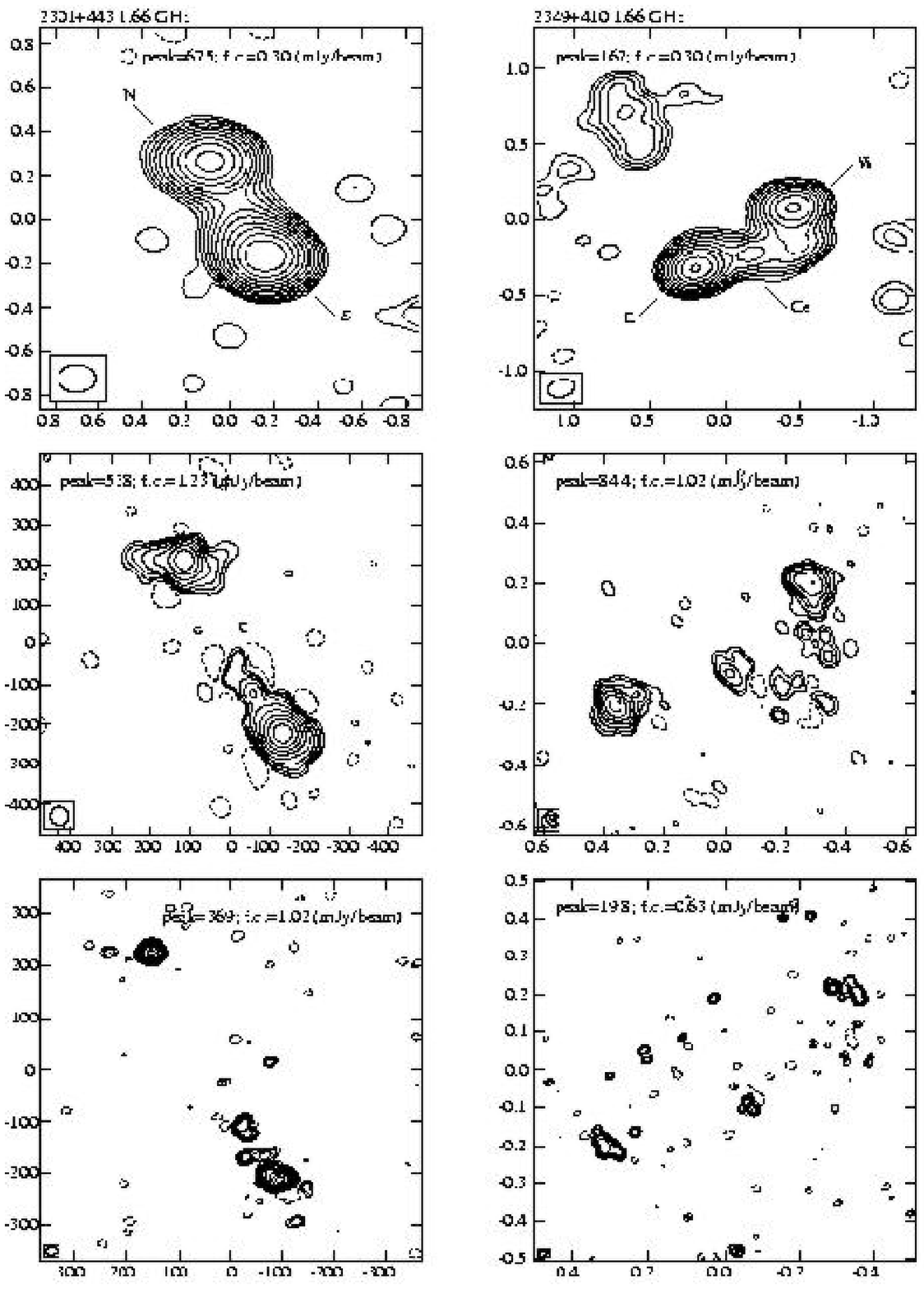}
}
\caption[]{{From top to bottom: MERLIN, combined EVN \& MERLIN and EVN
images (cont.)}}
\end{figure*}

\subsection {Source Parameters}
\begin{table*}
\begin{center}
\caption{Observational parameters of individual components}
\begin{tabular}
{lcr|rc|rc}
\hline
\hline
~~~Name &     c &  S$_{\rm M}$& S$_{\rm {E+M}}$& $\theta_1~~~~\theta_1$
~~~p.a&S$_{\rm {EVN}}$&$\theta_1~~~~\theta_1$ ~~~p.a\\
     &       &    mJy    &    mJy   &       mas~~~~~~    deg         &  mJy    &         mas~~~~~~     deg  \\   
(1)  & (2) & (3)  & (4) & ----- (5) -----  & (6) & ----- (7) -----  \\
\hline
\hline
0039+391&  W &   35   &   36  &  70$\times$39 ~~~34 &  [12] &                  \\
        &  C &        &    7  &  69$\times$18 ~~~34 &  ---  &                  \\
        &  E &  216   &  218  &  50$\times$45 ~~~98 &  143  & 44$\times$37~~105 \\
\hline
0110+401&  N &  373   &       &                     & [125] &                   \\
	&  S &   36   &   18  &  60$\times$31 ~~~~9 &   18  & 40$\times$15~~153 \\
        &ext & [78]   &  ---  &                     &  ---  &                   \\
\hline
0123+402&  N &   14   &  ---  &                     &  ---  &                   \\
        &  Ce&   55   &   36  &  69$\times$18 ~~~~3 &  [32] &                   \\
        &  C &        &    5  &      unres          &    6  &            unres  \\
        &  S &  146   &       &                     & [124] &                   \\
	&  S2&        &   75  &  92$\times$43 ~~160 &  ---  &                   \\
        &  S1&        &   70  &  73$\times$53 ~~174 &  ---  &                   \\
\hline
0140+387&   S&   22   &       &                     &  ---  &                   \\
        &Ce2 &   83   &   86  &  84$\times$47 ~~~47 &   25  & 36$\times$24~~~12 \\
	&Ce1 &   49   &   39  & 108$\times$43 ~~29 &  ---  &                   \\
        &   N&  176   &  150  &  66$\times$52 ~~~23 &   46  & 41$\times$14~~~10 \\
\hline
0255+460&  W &  396   &  313  &  94$\times$44 ~~115 & [128] &                   \\
        &  E &   94   &   74  &  87$\times$51 ~~120 &  ---  &                   \\
      &E tail&  44    &  [52] &                     &  ---  &                   \\
\hline
0722+393A& N &   807  &       &                     & [783] &                   \\
         & N2&        &  711  &  25$\times$15 ~~~72 &  744  & 25$\times$16~~~73 \\
         & N1&        &   35  &  47$\times$18 ~~104 &   37  & 16$\times$13~~114 \\
         & C &        &   --- &                     &    7  &  unres            \\
         & S &    73  &       &                     &  [49] &                   \\
	 &S2 &        &   56  &  88$\times$56 ~~~~2 &  ---  &                   \\
	 &S1 &        &   25  &    unres            &   24  & ~9$\times$5~~~~38 \\
\hline
0748+413B &  &   167  &   --- &                     &  ---  &                   \\
\hline
0754+396&N tail&[166] &       &                     &  [36] &                   \\
         & S &   260  &       &                     & [112] &                   \\
	 & S2&        &   78  &  95$\times$39 ~~~24 &  ---  &                   \\
	 & S1&        &   58  &  57$\times$28 ~~161 &  ---  &                   \\
\hline
0810+460B& S &   234  &       &                     &  ---  &                   \\
         & Ce&   253  &       &                     & [111] &                   \\
	 &   &        &   91  &  50$\times$35 ~~~48 &   36  & 27$\times$16~~~41 \\
         & N &   407  &       &                     & [161] &                   \\
\hline
0902+416& S  &   343  &       &                     &       &                   \\
        & S2 &        &  130  &  39$\times$22 ~~~12 &  105  & 29$\times$15~~~~4 \\
	& S1 &        &  214  &  36$\times$28 ~~~18 &  209  & 38$\times$25~~~~2 \\
        & C  &        &    4  &  21$\times$6 ~~~~~11 &    7  & 34$\times$16~~~49 \\
        & N  &    83  &       &                     &  [71] &                   \\
\hline
1027+392 & C &   182  &  ---  &                     & ---   &                   \\
        & ext&  [157] &  ---  &                     & ---   &                   \\
\hline
1039+424 & S &   146  &       &                     & ---   &                   \\
      &N tail&   [76] &  [55] &                     & ---   &                   \\
\hline
1128+455&  W &   142  & [123] &                     &  [59] &                   \\
        &  N &   589  & [533] &                     & [409] &                   \\
	& C  &        &   79  & 118$\times$18 ~135 &   22  & 42$\times$10~~143 \\
	& S  &  1034  &       &                     & [845] &                   \\
	&    &        &  866  &  55$\times$36 ~~178 &  312  & 31$\times$15~~161 \\
\hline
1157+460&  W &    83  &       &                     &  [52] &                   \\
        &    &        &   75  &  55$\times$55 ~~~50 &   37  & 37$\times$30~~~32 \\
        & S  &   368  &       &                     & [247] &                   \\
        &    &	      &  204  &  94$\times$44 ~~163 &   79  & 45$\times$21~~158 \\
	& N  &   474  &       &                     & [459] &                   \\
	&    &        &  386  &  35$\times$21 ~~131 &  315  & 26$\times$13~~127 \\
\hline
\hline
\end{tabular}
\label{comps}
\end{center}
\end{table*}

\setcounter{table}{1}
\begin{table*}
\begin{center}
\caption{Observational parameters (cont.)}
\begin{tabular}
{lcr|rc|rc}
\hline
\hline
~~~Name &     c &  S$_{\rm M}$& S$_{\rm {E+M}}$& $\theta_1~~~~\theta_1$
~~~p.a&S$_{\rm {EVN}}$&$\theta_1~~~~\theta_1$ ~~~p.a\\
     &       &    mJy    &    mJy   &       mas~~~~~~    deg         &  mJy    &         mas~~~~~~     deg  \\   
(1)  & (2) & (3)  & (4) & ----- (5) -----  & (6) & ----- (7) -----  \\
\hline
\hline
1212+380&  W &   126  &       &                     &  [33] &                   \\
        &  E &   121  &       &                     &  [61] &                   \\     
\hline
1241+411&  W &   34   &  [16] &                     &  ---  &                   \\
          &Ce&  220   &       &                     & [159] &                   \\
          &  &        &  164  &  25$\times$17 ~~~52 &  133  & 13$\times$6~~~~77 \\
          & E&   64   &       &                     &  ---  &                   \\
\hline
2301+443 & S &  725   &       &                     & [568] &                   \\
         &   &        &  662  &  26$\times$21 ~~~62 &  516  & 13$\times$10~~~85 \\
         &N  &  233   &       &                     & [142] &                   \\
         &   &        &  190  &  28$\times$24 ~~~75 &  136  & 14$\times$10~~~17 \\
\hline
2349+410 & W & [136]  &   93  &  88$\times$48 ~~~63 &  [31] &                   \\
	 &Ce &   23   &   16  &  76$\times$47 ~~~44 &  [14] &                   \\
	 &E  &  190   &       &                     &  [82] &                   \\
	 &   &        &  171  &  75$\times$38 ~~~49 &   72  & 41$\times$28~~~16 \\
\hline
\hline
\end{tabular}
\end{center}
\end{table*}

Source parameters are given in Table~\ref{comps} is as follows: 

\smallskip
\noindent
{\it 
Columns 1 and 2} --  Source name and sub--component label;

\smallskip
\noindent
{\it Column 3} --  MERLIN flux density (usually from JMFIT);

\smallskip
\noindent
{\it Column 4} --  EVN \& MERLIN flux density from JMFIT or TVSTAT
(in square brackets). Flux densities are not reported when they do not
differ significantly from MERLIN's. A ``---'' means no component visible 
at intermediate resolution;

\smallskip
\noindent
{\it Column 5} --  major and minimum, beam--deconvolved, angular size and major axis
position angle (from JMFIT) at EVN \& MERLIN resolution;

\smallskip
\noindent
{\it Columns 6 and 7} -- as for columns 4 \& 5 but for EVN--only images.

Source components are usually named from the lowest resolution image (MERLIN) 
as North ($N$), South ($S$), East ($E$), West ($W$), and Central ($Ce$). When 
a component is split into more pieces, a digit (1,2, etc) is added (e.g. 
$N1$, $N2$). Occasionally a candidate core ($C$) is mentioned. 

Depending on the component angular scale $vs$ resolution, flux densities are 
derived by  either
gaussian fits to the brightness  distribution (AIPS task JMFIT) or by
integration over the component brightness distribution (AIPS task TVSTAT,
values in square brackets in Table~\ref{comps}). In order not to give
redundant information we follow this approach:

-- for MERLIN--only images the flux densities of individual source components 
are always from JMFIT except for a few extended features for which we used TVSTAT 
(see Sect.~\ref {comm} for angular sizes);

-- at the higher resolutions, we give flux density, Half Maximum Widths (HMW) and 
p.a. (from JMFIT) for the brightest and most compact sub--structures only;

-- for the EVN--only images we give, in addition, also the component total flux  
density  ``seen'' by the interferometer (from TVSTAT, in square brackets), 
except when the latter is not too different from the value at lower
resolution;

-- for EVN \& MERLIN images this last information is generally unnecessary 
since the 
individual component total flux densities  usually agree with those ``seen'' by 
MERLIN. 

\begin{figure*}
\resizebox{\hsize}{!}{
\includegraphics{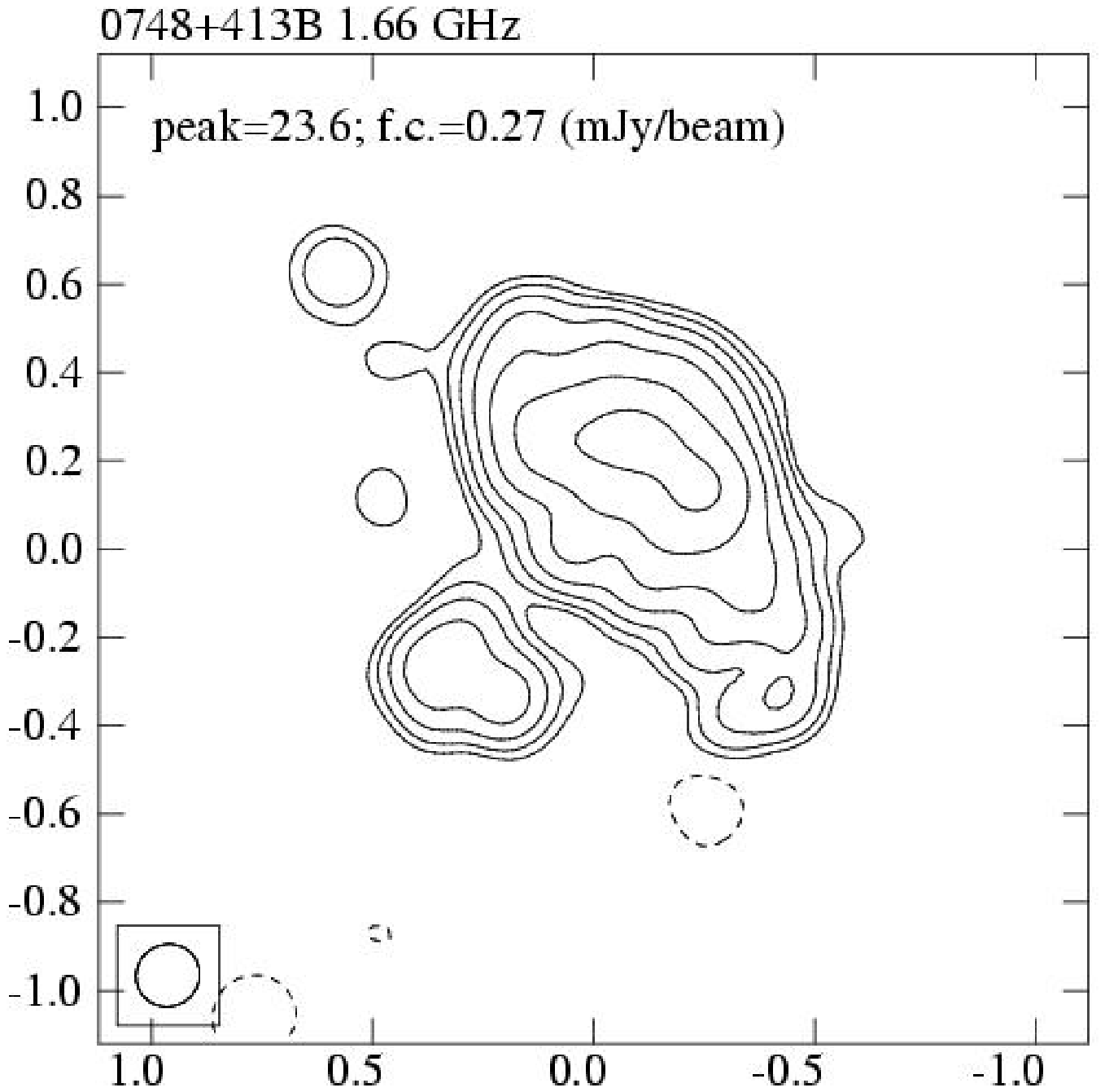}
\includegraphics{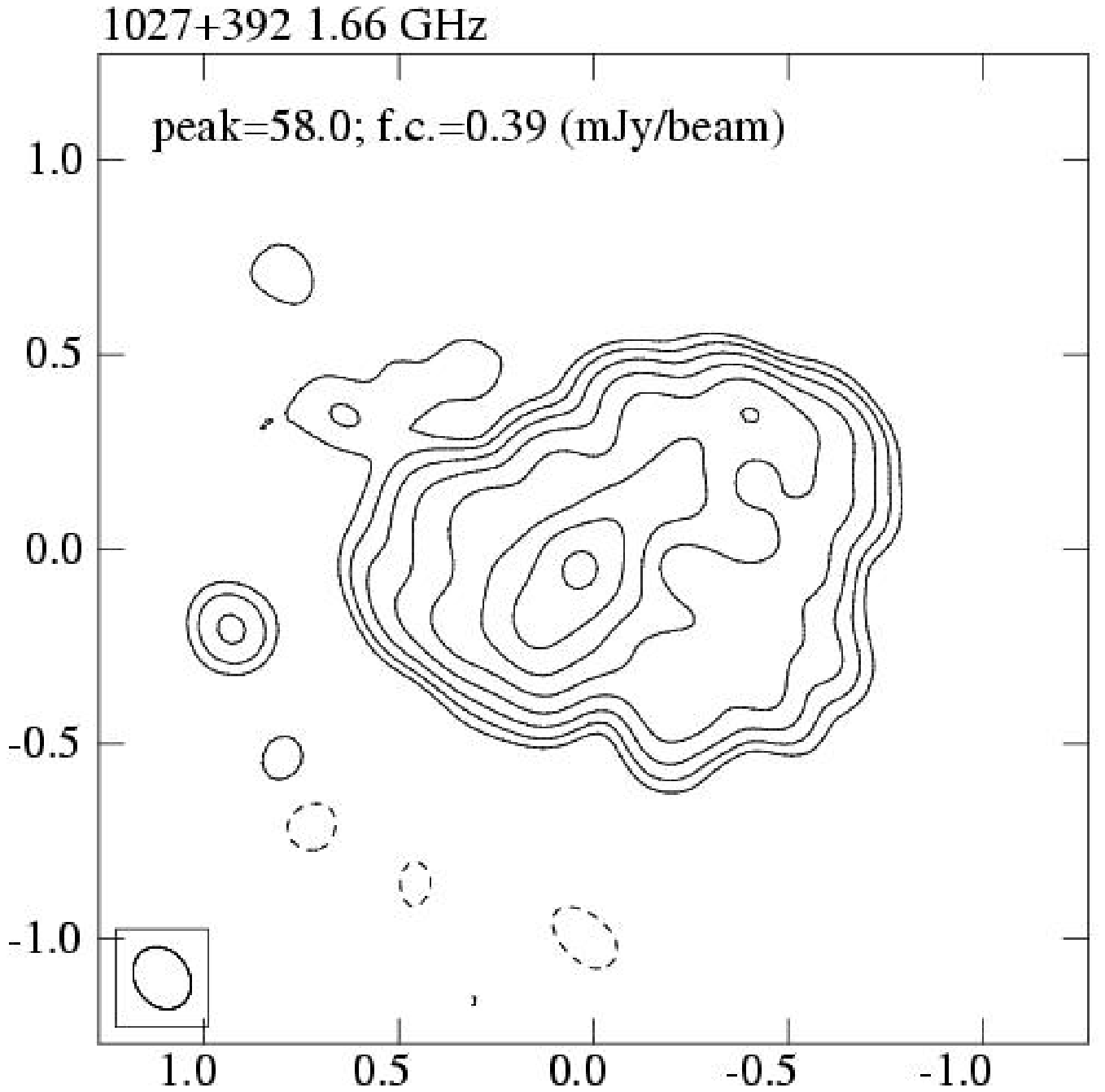}}
\caption[]{MERLIN--only images: the first contour (f.c.) is
three times the r.m.s. noise level on the image; contour levels
increase by a factor of 2; the restoring beam is shown in the bottom
left corner in each image.}
\label{mer-ima}
\end{figure*}

\section{Comments on individual sources}
\label{comm}
-- 0039+391: at MERLIN resolution this source is double, thus confirming
the hint of structure seen at 8.4 GHz in Paper I. At higher
resolutions a third component appears in between the two major ones. It
accounts for $\sim$2 mJy in the EVN image, and could host the source core, 
but we do not have spectral information to affirm this.

\smallskip
\noindent
-- 0110+401: part of the extended structure visible in the VLA images (Paper
I) is present in the MERLIN image as a two--sided tail, $\sim 2.2$ arcsec long, 
extending on both sides of the compact components. The total flux density in  
the combined image is $\approx$ 65 mJy (14 \%) lower than in the MERLIN one
since the extended structure is not visible here. 
In the EVN image several bright features are aligned along
an elongated bent structure, ending with a bright spot at each 
extreme. Structural uncertainties, however, are likely present due 
the low data quality of EVN (see Sect.~\ref{datar}).

\smallskip
\noindent
-- 0123+402: at MERLIN resolution the very asymmetric  triple structure
described in Paper I is confirmed, although all components show
now some substructure. At higher
resolutions component $N$ is mostly resolved out. This is likely the reason
why about 20 mJy (9 \%) of the MERLIN flux density is missing in the combined
image. Several bright knots appear in the other components, but 
the low quality of EVN data (see Sect.~\ref{datar}) reduces
the reliability of some of them. The EVN--only
image is not shown for this reason. The compact component $C$ could 
host or be the source core, although the lack of spectral information does not 
allow to confirm this. 

\smallskip
\noindent
-- 0140+387: in Paper I this source looks like a simple double, while at
MERLIN and EVN \& MERLIN resolutions the two main components appear connected 
by a bridge of
emission. All the components are resolved in the combined image.
In the EVN image part of the extended emission is still present, although 
totally fragmented. Only the two extreme spots stand out and are reliably
detected, while no unresolved component is visible to suggest the presence of
the candidate core.

\smallskip
\noindent
-- 0255+460: the double structure seen in Paper I is confirmed
but the Eastern component displays a sort of curved tail, $\sim 0.8$ arcsec in
size, pointing to North and
then to East (E tail in Table~\ref{comps}). About 12 \% of the total flux
density ($\approx$ 70  mJy) is missing in the MERLIN image,
probably in the extended eastern component. In the EVN image only
component $W$ is clearly
visible: it appears as a wiggling structure accounting for 128 mJy ($\approx
85$ \% of the EVN total flux density). Some weak low brightness emission is
present at the position of component $E$.

\smallskip
\noindent
-- 0722+393A: at MERLIN resolution the source appears as a very asymmetric
double (component flux density ratio 11:1), as it could have been already
guessed (Paper I). At intermediate resolution more
structure is present. 
In the EVN image, a number of knots are visible. Remarkable are the hot--spot 
$S1$ and component $C$, which, due to its location, could be the source core. 
However no spectral information
is available to confirm this. The confusing source B3 0722+393B of $\sim 200$  
mJy at 1.4 GHz (FIRST, Becker et al. \cite{beck}) $\sim 6.3$ arcmin to
South--East had to be subtracted from the MERLIN visibility data.

\smallskip
\noindent
-- 0748+413B: only the MERLIN image (Fig.~\ref{mer-ima}) could be produced 
(see Sect.~\ref{s-ima}).
The source shape could recall the ending section of a lobe whose core is 
located somewhere to the South--East.
Actually a weak source ($S_{1.5}$=17.3 mJy, Paper I) is present 13.8 arcsec in
this direction. No emission is visible  between the two structures neither 
on FIRST (Becker et al. \cite{beck}) nor in a low resolution MERLIN image
(not shown). About 40 mJy (19 \%) are missing 
with respect to the expected total flux density, only part of which due to the 
secondary component.

\smallskip
\noindent
-- 0754+396: the $\sim 2.6$ arcsec long structure visible in
Paper I is confirmed in great detail in the MERLIN image, although some 30 mJy (
7 \%) are missing. In the EVN \& MERLIN
image the strong $S$ component starts with a bright hot--spot ($S1$) and
extends to North to join a wiggling fragmented jet. The EVN--only image (not shown) presents only a set of
knots whose location is rather uncertain given the low quality of EVN data
(see Sect.~\ref{datar}). This source was classified as ``cJ?'' in
Paper I but a ``standard'' clear 
core component is not visible. The EVN \& MERLIN image could also be 
reminiscent of a Narrow Angle Tail
($NAT$) radio galaxy, as often seen in galaxy clusters, usually on larger scales.
A weak source of $S_{1.5}$=5.5 mJy,  11.1 arcsec 
to South West (Paper I) is likely an unrelated source since no emission 
is visible between the two structures neither on FIRST (Becker et al.
\cite{beck}) nor in a low resolution 
MERLIN image (not shown). 

\smallskip
\noindent
-- 0810+460B: the double structure seen in Paper I is now
split into three well resolved components, none of which has a flat spectrum.
When the EVN \& MERLIN image is displayed in colours, component $S$ appears  
as a twisted tail. In the EVN image features $N$ and $Ce$ are well resolved 
and look like misaligned lobes while component $S$ is totally resolved out. 

\smallskip
\noindent
-- 0902+416: at the resolution of MERLIN the source is an asymmetric double, the
southern component accounting for 80 \% of the source total flux density.
The EVN image confirms this basic morphology, displaying additional structure.
The weak central component $C$ could host the source core, but
we have no spectral information to confirm this. 

\smallskip
\noindent
-- 1027+392: the MERLIN-only image (Fig.~\ref{mer-ima}) 
confirms the structure described in Paper I,
with some more detail, but it does not have enough resolution to 
understand the structure of this object any better. A possibility could be
that we are seeing only one lobe of a double source extremely asymmetric in
flux density like, for instance, 3C299  (van Breugel et al. \cite{vB}) 
whose secondary component, 12 arcsec away, accounts at 1.7 GHz for only 
$\approx$ 5 \% of the main 
component flux density. However no other feature is visible on the FIRST
(Becker et al. \cite{beck}) images within 
a few arcmin at a dynamic range level $~\sim 1000:1$.
No EVN information is available on this source (see Sect.~\ref{datar}).

\smallskip
\noindent
-- 1039+424: in Paper I the source structure was described as a core--jet.
The EVN \& MERLIN image, instead, rather resembles either a Narrow Angle Tail
($NAT$) radio source, with a tail  $\sim 2$ arcsec long pointing to North,
or, alternatively, one lobe (the southern one?) of a double radio
source of which we do not detect the companion. A weak source of
S$_{1.5}$=5.9 mJy is present 15.5 arcsec North (Paper I) but no
emission is visible between the two structures neither on FIRST
(Becker et al. \cite{beck}) nor in a low resolution MERLIN image 
(not shown).
About 20 mJy (9 \%) of the MERLIN flux density are missing in the
combined image. The EVN observations allow to barely detect only the
compact feature, accounting for 8 mJy, seen embedded in the extended
$S$ lobe. 

\smallskip
\noindent
-- 1128+455: the MERLIN image shows a very distorted triple source, as in
Paper I, with component $S$ accounting for $\sim$60 \%
of the source total flux density. All components have a steep spectrum, thus
excluding the possibility that any of them is the source core. At EVN
resolution component $W$ is marginally detected, while a weak
feature ($C$) shows up in between features $N$ and $S$. We mark it in
Fig.~\ref{images} as the possible core, although there is no spectral
information to confirm this. A possibility is that we are in  the
presence of a small Wide Angle Tail ($WAT$) radio galaxy, as often seen
in galaxy clusters, usually on larger scales, with the  center of the
host galaxy located at component~$C$. 

\smallskip
\noindent
-- 1157+460: the MERLIN image displays a peculiar triple source,
whose components form an angle of about 90$^\circ$, confirming the structure
seen in Paper I. In the EVN image weak hot--spots are seen
at the edges of components $N$ and $W$, from where weak
tails emerge pointing to component $S$. One cannot exclude, however, that
component $W$ be an unrelated source.

\smallskip
\noindent
-- 1212+380: the MERLIN image resolves the structure shown in Paper I
into a double radio source, symmetric in flux density but not
in shape, component $E$ being more elongated. 
At EVN resolution structural uncertainties are likely present due 
the low data quality of EVN (see Sect.~\ref{datar}).

\smallskip
\noindent
-- 1241+411: the MERLIN image shows a triple structure, with misoriented
components. Component $W$, barely detected in the EVN \& MERLIN image,
disappears completely  in the EVN image. At EVN resolution a bright compact 
component, accounting for $\approx$ 95 \% of the EVN total flux density, shows 
up at $Ce$. Also this source could be a small
Wide Angle Tail ($WAT$) radio galaxy, as often seen in galaxy clusters, usually on 
larger scales, with the  center of the host galaxy located at component $Ce$.
The confusing source B3 1242+410 $\sim 6.5$ arcmin to
South--East, of $\sim 1.37$ Jy at 1.4 GHz (FIRST, Becker et al. \cite{beck})  had to be
subtracted from the MERLIN visibility data.

\smallskip
\noindent
-- 2301+443: in the MERLIN image about 8 \% ($\approx$ 90mJy) of the total flux 
density is missing. Both components are mostly resolved in the EVN
image which shows only some hints of extended structure
along the source axis and two bright spots.
They could represent the hot--spots of the source. In the EVN \&MERLIN more 
extended structure is visible in the form of two misoriented lobes.

\smallskip
\noindent
-- 2349+410: in Paper I the source shows a quite strange
triple structure, but in the MERLIN image the northern feature (accounting for
21 mJy at 1.66 GHz) is detached from
the rest of the structure, suggesting an independent source.
The two other components ($W$ and $E$) are connected by a sort of bridge,
where a weaker knot ($Ce$)
is also visible. At EVN resolution only the most compact sub--structures are
present. The central knot, accounting for $\sim$ 14
mJy, could host the source core,
although we do not have any spectral information to affirm this. 
About 50~mJy ($\sim$15 \%) are missing in the combined image.

\section {Discussion}

The MERLIN observations, at a resolution roughly a factor of two
better than that of the VLA at 8.4 GHz, 
basically confirm the known morphology. The EVN \& MERLIN and 
EVN--only observations have produced images with high enough
resolution to reveal many more details in the source morphology for the
large majority of the sources.

At MERLIN resolution most sources (13 out of 18) show a double or triple,
sometimes quite distorted, structure. The high resolution images also show 
that they are preferentially edge brightened. 
Furthermore we see, in a minority of objects, a compact component,
centrally located, which
could be the source core. However having only one, relatively low, frequency
we restrain ourselves from making any firm statement about.

In spite of the lack of strong evidence of core detection,
the overall morphology indicates that most sources are two--sided. We
are convinced that we see lobes, often connected with bridges or, perhaps, 
jets.
We classify these sources as MSOs (Table~\ref{tabsample}).
We note that in this sample
the lobes tend to be quite asymmetric in flux density. It has been already
remarked  that in CSOs and MSOs large lobe flux density asymmetries
sometime do exist, but here the situation seems more extreme. For instance 
in six
out of the 13 possible MSOs the flux density ratio of the two lobes is larger than three.
It is very likely, however, that this
is the result of the performed source selection from the B3-VLA CSS sample.
In a number of the MSOs the lobes show low brightness wings or tails
often distorted from the main source axis, not dissimilar from what found
in large size powerful radio sources.

We find it difficult to unambiguously classify five sources (0110+401,
0748+413B, 0754+396, 1027+392 and 1039+424).

Further, in two cases ($1157+460$ and $2349+410$) another source is
found close in projection to the main one, likely unrelated to
it. Four other sources ($0754+396$, $1039+424$,  $1128+255$ and
$1241+411$) could be alternatively classified as $NAT$ or $WAT$. 
It is tempting to speculate on the presence of a galaxy cluster around
these six radio sources. 

We have computed the equipartition parameters for the source
components, under the following assumptions: a) proton to electron
energy ratio of one; b) filling factor of one; c) maximum and minimum
electron (and proton) energies  corresponding to synchrotron emission
frequencies of 100 GHz and 10 MHz; d) ellipsoidal volumes with axis
corresponding to the observed ones. We also computed the brightness
temperatures ($T_{\rm B}$) of each component.

In order to make easier the reading of the paper, we do not report the 
individual component values, since they can be obtained from the parameters in 
Table~\ref{comps}, but only mention the typical values.

At the resolution of MERLIN--only images, we find:

\noindent
{\it i)} equipartition
magnetic field $H_{\rm eq}\ltsim 1.5$ mG,  up to $\approx~3$~mG; {\it ii)}
energy densities $u_{\rm min}\ltsim 1.5\times 10^{-7}$ erg/cm$^3$, up to $\sim 
10^{-8}$ erg/cm$^3$; {\it iii)} $T_{\rm B}\ltsim 6\times10^7$~K, up to $\sim
4\times10^9$~K. 

The most compact sub--components detected at the higher EVN resolution
have, on average, equipartition magnetic fields higher 
by roughly a factor of two and energy densities and brightness temperatures higher
by about a factor of four.

\section{Conclusions }

We have presented the results of EVN and  MERLIN observations of a 
sub--sample of 18 radio sources from the \\ B3--VLBA CSS sample of Paper I,
which were just marginally resolved with the VLA.

The majority of the sources are classified as MSOs, a number
of which very asymmetric in component flux density. At difference of what
found in Paper II, no very 
bright compact component has been detected. This may be due to the lower 
resolving power of the intra--European EVN configuration used
in the present work compared to that of VLBA, but could also be
intrinsic, since the radio sources presented in this paper are a few
times more extended than those presented in Paper II.

Equipartition parameters are consistent with other findings (e.g. Paper II)
when allowance is made for the larger component sizes. In particular we note 
that brightness temperatures hardly reach $10^9$ K, some hundred times lower 
than those of the sources of Paper II.
Together with the VLA data on the large size sources and with the VLBA 
data presented in Paper II on the smallest sources,
we have a new determination of the Linear Size distribution
in the range $\approx 0.2 h^{-1} < LS$(kpc) $<$ 20 $h^{-1}$. 

This will be discussed it in a separate paper.
\begin{acknowledgements}
MERLIN  is the Multi--Element Radio Linked Interferometer Network and is a
national facility operated by the University of Manchester on behalf of
PPARC. The European VLBI Network is a joint facility of European and Chinese
Radio Astronomy Institutes funded by their National Research Councils.
This work has been partially supported by the Italian MURST under
grant COFIN-2001-02-8773.
\end{acknowledgements}

\end{document}